\documentstyle[psfig]{article}
\begin{document}
\pagenumbering{arabic}

\title{Thin Layer Approximation in 3-D}

\author{
 G. S. Bisnovatyi-Kogan
 \thanks{ Space Research Institute, 84/32 Profsoyuznaya, Moscow,
 117810, Russia}
 \and
 S. A. Silich
 \thanks
 {Main Astronomical Observatory of the National Academy of
 Sciences of Ukraine 252650, Kiev, Golosiiv, Ukraine}}
\maketitle

\begin{abstract}
Equations are derived which describe a propagation of strong shocks in the
interstellar matter , without any demands for a symmetry, in a thin
layer approximation (2.5 dimensions). Using these equations permits
to calculate a propagation of shock waves from nonsymmetric supernovae
explosions in the medium with arbitrary density distribuion,
formation of superbubbles in galaxies.
\end{abstract}

\section {Thin layer approximation}

A thin shell approximation for discription of strong shocks
is based on two simplifications. First,
it is assumed that all swept-up intercloud gas accumulates into the
thin shell just behind the shock front and moves with the velocity
$u$. Second,
the pressure distribution inside the cavity $P(r,t)$ is taken to be uniform.
The equations of mass and momentum conservation may be expressed as
follows (Cherny, 1957) :
\begin{equation}
      \label{sg1.36}
M=M_0 + 4\pi \int_{0}^R {\rho}(r)r^2dr,\quad
\frac{d(Mu)}{dt} = 4{\pi}R^2(P_{\rm in}-P) + Mg,
\end{equation}
where $M$ is the mass of the shell, $M_0$ is the ejected mass, $R$ is the
shock radius and $u$ is the gas velocity behind the shock;
$\rho (r)$ and $P$ are the density and the
pressure of the ambient gas, $g$ is the external gravitational field.
For the adiabatic blastwave without gravity
 \begin{equation}
      \label{sg1.37}
1)\,\frac{dR}{dt}=\frac{\gamma + 1}{2}u \quad {\rm or} \quad
2)\, \frac{dR}{dt}=u; \qquad
 E_0= E_{\rm th} + \frac {1}{2} M u^2,
\end{equation}
where $E_0={\rm const}$ is the energy of the explosion,
$E_{\rm th}=\frac{4\pi}
{3(\gamma -1)}P_{\rm in}R^3$ is the thermal energy of the blastwave, and
$\gamma$  is the adiabatic power.
Equations (~\ref{sg1.36})-(~\ref{sg1.37}) have a simple solution for the
homogeneous case if the swept-up mass is much
greater than the ejected one
$R=\left[ \frac{\xi_0 E_0}{\rho_0} \right]^{1/5}t^{2/5}$,
where (Chernyi, 1957, Bisnovatyi-Kogan and Blinnikov, 1982)

 \begin{equation}
       \label{sg1.41}
1)\, {\xi}_0=\frac {75(\gamma-1)(\gamma+1)^2}{16 \pi (3 \gamma -1)}, \quad
2)\,\xi_0 = \left[\frac{75(\gamma - 1)}{8 \pi} \right]^{1/5}.
\end{equation}
Comparison with an exact self-similar (SS) solution (Sedov, 1946) shows, that
the case 2) gives much better precision, $\xi_0^{1/5}=1.033,\,\,1.014 \,\,
1.036$ for (SS), (1), (2) cases respectively at $\gamma=1.4$; and similarily
$\xi_0^{1/5}=1.15,\,\,1.12,\,\,1.15$ for $\gamma=5/3$.

\section{Three-dimensional shocks}

Introduce following Bisnovatyi-Kogan and Silich (1991), Silich (1992)
(see also Palou\v{s}, 1990, Bisnovatyi-Kogan and Silich, 1995)
Cartesian coordinate system $(x,y,z)$. Let $m$ is the mass, $\bf r$ is the
radius-vector, $\bf u$ is the velocity of the particular Lagrangian
element of the shock,
$\rho (x,y,z) = \rho_0 f(x,y,z)$ is the ambient gas density,
${\bf n}$ is the normal to the shock front unit vector, ${\bf g}$
is the
acceleration of the external gravitational field, ${\bf V}$ is the velocity
field of the undisturbed gas flow, $\Sigma$ is the surface area of the
Lagrangian element, $m = \sigma \Sigma$, $\sigma$ is the surface density,
$\Delta P = P_{\rm in} - P$ is the pressure difference between the
hot interior and warm (cold) external gas. The pressure
$P_{\rm in} = (\gamma - 1) E_{\rm th} / \Omega$
of the hot tenuous gas within the cavity is a function of the bubble
thermal energy $E_{\rm th}$ and volume $\Omega$.
To describe shock expansion we must introduce element of the surface area and
define volume of any closed three-dimensional region. It is well known from
the differential geometry that any surface may be specified parametrically,
with the Cartesian coordinates at any point on the surface being function
of two parameters, $\lambda_1$ and $\lambda_2$: $x=x(\lambda_1,\lambda_2)$,
$y=y(\lambda_1,\lambda_2)$, $z=z(\lambda_1,\lambda_2)$. Then the element of
the surface area may be defined by the expression (Budak and Fomin, 1965)

\begin{eqnarray}
 & &    \hspace{-0.1cm}
        \label{bs3.5}
d \Sigma = S(\lambda_1,\lambda_2) d\lambda_1 d\lambda_2 ,
        \\ [+.15cm]
 & &     \hspace{-0.1cm}
        \label{bs3.6}
 S(\lambda_1,\lambda_2) =
\left\{\left[ \frac{\partial(y,z)}{\partial
 (\lambda_1,\lambda_2)} \right]^2 +
 \left[ \frac{\partial(z,x)}{\partial(\lambda_1,\lambda_2)} \right]^2 +
 \left[ \frac{\partial(x,y)}{\partial(\lambda_1,\lambda_2)}
 \right]^2\right\}^{1/2},
\end{eqnarray}
where $\partial(x_i,x_j) / \partial(\lambda_1,\lambda_2) \equiv$
$det(\partial (x_i,x_j) / \partial (\lambda_1,\lambda_2))$ are the
ap\-pro\-pri\-ate Ja\-co\-bi\-ans.
Similarly could be obtained expressions for $\Omega$ and ${\bf n}$

If parameters $\lambda_1$ and $\lambda_2$ are considered as the Lagrangian
coordinates of the shock front, then the equations for 3-D shock propagation
may be written for the mass $\mu = \sigma S(\lambda_1,
\lambda_2)$ per unit of Lagrangian square, in a compact form,
convenient for numerical integration

\begin{eqnarray}
 & &    \hspace{-0.5cm}
        \label{bs3.9}
\frac{{\rm d} \mu}{{\rm d}t} = \rho \chi,
        \\ [+.15cm]
 & &     \hspace{-0.5cm}
         \label{bs3.10}
\frac{{\rm d} u_x}{{\rm d}t} = \frac{\Delta P}{\mu}
\frac{\partial (y,z)}{\partial (\lambda_1 \lambda_2)} - \frac{u_x-V_x}{\mu}
     \rho \chi + g_x,
        \\ [+.15cm]
 & &     \hspace{-0.5cm}
         \label{bs3.11}
\frac{{\rm d} u_y}{{\rm d}t} = \frac{\Delta P}{\mu}
\frac{\partial (z,x)}{\partial (\lambda_1 \lambda_2)} - \frac{u_y-V_y}{\mu}
     \rho \chi + g_y,
        \\ [+.15cm]
 & &     \hspace{-0.5cm}
         \label{bs3.12}
\frac{{\rm d} u_z}{{\rm d}t} = \frac{\Delta P}{\mu}
\frac{\partial (x,y)}{\partial (\lambda_1 \lambda_2)} - \frac{u_z-V_z}{\mu}
     \rho \chi + g_z,
        \\ [+.15cm]
 & &     \hspace{-0.5cm}
         \label{bs3.13}
\frac{{\rm d}x}{{\rm d}t}=u_x, \quad \frac{{\rm d}y}{{\rm d}t}=u_y, \quad
\frac{{\rm d}z}{{\rm d}t}=u_z.
\end{eqnarray}
Here function $\chi$ is defined as follows:

\begin{equation}
      \label{bs3.14}
\chi = (u_x-V_x)\frac{\partial(y,z)}{\partial (\lambda_1,\lambda_2)} +
 (u_y-V_y)\frac{\partial(z,x)}{\partial(\lambda_1,\lambda_2)} +
 (U_z-V_z)\frac{\partial(x,y)}{\partial(\lambda_1,\lambda_2)}.
\end{equation}
Approximating the shock
front by a number $N$ of Lagrangian elements one get a system of $7N$
differential equations of the mass and momentum conservation. This set of
equations is coupled with the equation for the gas pressure
within the cavity, and the equation of energy for
$E_{\rm tot} = E_{\rm th} + E_{\rm k} + E_{\rm g}$, consisting from
the thermal energy of the hot bubble
interior, kinetic and gravitational energies of the shell.
The kinetic and gravitational energies of the shell
are determined by corresponding surface integrals.
Variations of
the total energy $E_{\rm tot}$ of the remnant throughout the adiabatic stage
of evolution are defined by the energy input rate $L(t)$, kinetic and
thermal energies of the swept-up interstellar gas with temperature $T(x,y,z)$:

\begin{equation}
      \label{bs3.16}
E_{\rm tot} = E_0 + \int_0^t [ L(t)  +
\frac{1}{2}
\int_{\lambda_{1,min}}^{\lambda_{1,max}} \int_{\lambda_{2,min}}^
{\lambda_{2,max}}
\dot \mu (V^2+3kT/\eta) {\rm d} \lambda_1 {\rm d} \lambda_2 ] {\rm d}t,
\end{equation}
where $E_0$ is the initially deposited energy, $\eta$ is the mean mass per
particle.
On the radiative phase of expansion the gas behind the shock front cools so
quickly that does not add to the total energy of the remnant. Rarefied hot
gas inside the cavity expands adiabatically and accelerates the surrounding
dense shell. Time-derivative of the thermal energy of the remnant is
defined then by the equation, which is used instead of (~\ref{bs3.16}):

\begin{equation}
      \label{bs3.19}
\frac{dE_{\rm th}}{dt} = L(t) - \int_{\lambda_{1,min}}^{\lambda_{1,max}}
\int_{\lambda_{2,min}}^{\lambda_{2,max}}
 P_{\rm in} u_n S(\lambda_1,\lambda_2) {\rm d}\lambda_1 {\rm d}\lambda_2 ,
\end{equation}
where $u_n$ is the velocity component normal to the shock front.
Calculations of supershell formation in the plane-stratified
differentially rotating Galactic disk have been performed by
Silich {\it et al.} (1994) and  are represented in the
Fig.\ \ref{fig23}.
An hourglass remnant with a noticeable degree
of deformation by the Galactic shear have appeared.
Formation of elongated superbubbles due to differential galactic rotation
gives a possibility to determite unambiguously a direction
of the galactic rotation (Mashchenko and Silich, 1995). The means of this
determination is shown in the Fig.\ \ref{fig31}.

\begin{figure}[thbp]
\centerline{
\psdraftbox
}
\caption[]{Galactic superbubble morphology for different locations
of the parent OB-association relative to the galactic plane. Left: the
OB-association is at the midplane of the Galaxy. Right: the OB-association is
50pc above the Galactic plane.}
\label{fig23}
\end{figure}

\begin{figure}[thbp]
\centerline{
\psdraftbox
}
\caption[]{Scheme of the HI holes orientation in the plane of view.}
\label{fig31}
\end{figure}


\begin{thebibliography}{}

\bibitem{bkb82}
Bisnovatyi-Kogan G. S. and Blinnikov S.I. 1982, Astron.
Zh. ${\bf 59}$, 876 [Sov. Astron. ${\bf 26}$, 530 (1982)].

\bibitem{bks91}
Bisnovatyi-Kogan, G. S. and Silich S.A., 1991, Astron. Zh.
${\bf 68}$, 749 [Sov. Astron. ${\bf 35}$, 370 (1991)].

\bibitem{bks95}
Bisnovatyi-Kogan, G. S. and Silich S.A., 1995, Rev. Mod. Phys.
${\bf 67}$, 661.

\bibitem{bf67}
Budak B. M. and Fomin S.V. 1967, Multiple integrals and series
(Nauka, Moscow).

\bibitem{c57}
Chernyi G. G., 1957, Doklady Acad. Sci. USSR ${\bf 112}$,
213.

\bibitem{ms95}
Mashchenko S. Ya. and Silich S.A.  1995, Astron. Zh.
${\bf 72}$, 660.

\bibitem{p90}
Palou\v{s} J., 1990, in The Interstellar Disk-Halo
Connection in Galaxies, edited by H. Bloemen (Srerrewacht, Leiden)
p. 101.

\bibitem{s46}
Sedov L. P., 1946, Dok. Akad. Nauk SSSR, ${\bf 42}$, 17.

\bibitem{s92}
Silich, S. A., 1992, Astrophys. Space. Sci, ${\bf 195}$, 317.

\bibitem{sfp94}
Silich S. A., Franco J. Palou\v{s}, J. and
Tenorio-Tagle G. 1994, in Violent Star Formation from 30 Doradus to QSOs,
edited by G. Tenorio-Tagle (Cambridge Univ. Press, Cambridge) p. 162.

\end{thebibliography}
\end{document}